\documentclass{aa} 

\usepackage{graphics}

\begin{document} 
 
\thesaurus{03     %
           (13.07.1)} 
 
 
\title{The broken light curves of gamma-ray bursts GRB~990123 and GRB~990510} 
 
\author{S. Holland\inst{1} 
           \and 
        G. Bj{\"o}rnsson\inst{2} 
           \and 
        J. Hjorth\inst{3} 
           \and 
        B. Thomsen\inst{1} 
       } 
 
\offprints{Stephen Holland} 
 
\institute{Institut for Fysik og Astronomi (IFA), 
           Aarhus Universitet, 
           Ny Munkegade, Bygning 520, 
           DK--8000 {\AA}rhus C, 
           Denmark
           e-mail: holland@ifa.au.dk, bt@ifa.au.dk 
         \and 
           Science Institute, 
           University of Iceland, 
           Dunhaga 3, 
           IS-107 Reykjavik, Iceland 
           e-mail: gulli@raunvis.hi.is 
         \and 
           Astronomical Observatory, 
           University of Copenhagen, 
           Juliane Maries Vej 30, 
           DK--2100 Copenhagen {\O}, 
           Denmark
           e-mail: jens@astro.ku.dk 
          } 
 
\date{received 9 June 2000; accepted} 

\titlerunning{GRB~990123 and GRB~990510}
 
\maketitle 

 
 
\begin{abstract} 

	We have collected all of the published photometry for
\object{GRB~990123} and \object{GRB~990510}, the first two gamma-ray
bursts where breaks were seen in the light curves of their optical
afterglows, and determined the shapes of their light curves and the
break times.  These parameters were used to investigate the physical
mechanisms responsible for the breaks and the nature of the ambient
medium that the bursts occurred in.  The light curve for
\object{GRB~990123} is best fit by a broken power law with a break
$1.68 \pm 0.19$ days after the burst, a slope of $\alpha_1 = -1.12 \pm
0.08$ before the break, and a slope of $\alpha_2 = -1.69 \pm 0.06$
after the break.  This is consistent with a collimated outflow with a
fixed opening angle of $\theta_0 \approx 5\degr$.  In this case the
break in the light curve is due to the relativistic fireball slowing
to $\Gamma \approx 1/\theta_0$.  The light curve for
\object{GRB~990510} is best fit by a continuous function with an
early-time slope of $\alpha_1 = -0.54 \pm 0.14$, a late-time slope of
$\alpha_2 = -1.98 \pm 0.19$, and a slow transition between the two
regimes approximately one day after the burst.  This is consistent
with a collimated outflow with $\theta_0 \approx 5\degr$ that is
initially radiative, but undergoes a sideways expansion that begins
approximately one day after the burst.  This sideways expansion is
responsible for the slow break in the light curve.

\keywords{Gamma rays: bursts}
 
\end{abstract} 
 
 
\section{Introduction\label{SECTION:intro}} 
 
        In the first few months of 1999 bright optical afterglows
(OAs) were observed for two gamma-ray bursts (GRBs):
\object{GRB~990123} and \object{GRB~990510}.  Extensive follow-up
observations were made at optical wavelengths for both GRBs, so it was
possible to follow their light curves down to $V \approx 28$, with the
deepest photometry being done with the \emph{Hubble Space Telescope}
(\emph{HST}).  These data, when put onto a common photometric system,
can be used to constrain the shape of each burst's light curve.  The
resulting shape parameters (the time of the break, and the slopes
before and after the break) can be used to constrain physical models
of the OA, and the interaction of the relativistic ejecta from the
initial blast with the surrounding medium.
 
        At 9:46:56.12 UT on 1999 January 23 the Burst and Transient
Source Experiment (BATSE) detectors on the \emph{Compton Gamma Ray
Observatory} satellite detected \object{GRB~990123} in the
constellation Bo{\"o}tes.  $X$-ray observations from the
Dutch--Italian satellite \emph{BeppoSAX} localized the position of the
burst to within $5\arcmin$ (Piro~\cite{P99}) and the Robotic Optical
Transient Search Experiment (Akerlof \& McKay~\cite{AM99}; Akerlof
et~al.~\cite{ABB1999}) detected an OA that reached a peak magnitude of
$V = 8.86 \pm 0.02$ just 47 seconds after the BATSE trigger.  Further
optical observations constrained the redshift to between $1.60 \le z <
2.04$ (Andersen et~al.~\cite{AC99}; Kulkarni et~al.~\cite{KD99}), and
images taken with the \emph{HST} Space Telescope Imaging Spectrograph
(STIS) revealed a host galaxy with $V_0 = 24.25$ (Holland \&
Hjorth~\cite{HH99}).  The burst had a peak fluence of $(5.09 \pm 0.02)
\times 10^{-4}$ erg cm$^{-2}$ (Kippen~\cite{K99a}), which places it in
the top 0.3\% of the BATSE fluence distribution.  This corresponds to
a total energy release, assuming isotropy and depending on the details
of the adopted cosmology, of between $\approx 3 \times 10^{54}$ and
$\approx 4.5 \times 10^{54}$ erg, which is approximately the rest-mass
energy of two neutron stars.  This amount of energy is difficult to
explain using merging compact objects since it would require that the
entire rest mass of both objects be converted to gamma radiation with
an efficiency of nearly 100\%.
 
        \object{GRB~990510} (the Anja burst) was detected by both the
\emph{Compton Gamma Ray Observatory} and \emph{BeppoSAX} in the
constellation Octans at 8:49:05.95 UT on 1999 May 10.  Optical
follow-up observations 3.5 hours after the burst revealed an OA with
$R \approx 17.5$ (Axelrod et~al.~\cite{AMS99}).  Vreeswijk
et~al.~(\cite{VG1999}) found a lower limit for the redshift of $z =
1.619$.  A faint galaxy ($V = 28.5 \pm 0.5$) has been detected
$0\farcs066 \pm 0\farcs009$ from the GRB (Bloom~\cite{B2000}; Fruchter
et~al~\cite{FHP2000}).  The burst had a peak fluence of $(2.56 \pm
0.09) \times 10^{-5}$ erg cm$^{-2}$ (Kippen~\cite{K99b}), which places
it in the top 9\% of the BATSE fluence distribution.  This corresponds
to a total isotropic energy release, depending on the details of the
cosmology, of at least $10^{53}$ erg, approximately one order of
magnitude less than the energy released in \object{GRB~990123}.
 
        The light curve for \object{GRB~990123} became steeper
approximately two days after the burst (Castro-Tirado
et~al.~\cite{CT99}, Kulkarni et~al.~\cite{KD99}), while Harrison
et~al.~(\cite{HB99}), Israel et~al.~(\cite{IMC1999}), and Stanek
et~al.~(\cite{SGK1999}) found that the light curve of
\object{GRB~990510} became significantly steeper approximately 1.5
days after the burst.  These breaks have been interpreted as evidence
that the outflows from the bursts are collimated with opening angles
of approximately $5\degr$--$10\degr$ (Sari et~al.~\cite{SP99},
Castro-Tirado et~al.~\cite{CT99}).  Such models provide a solution to
the so-called ``energy crisis'' of GRBs since, if GRBs are collimated
outflows, the total energy requirement drops by factor of between
approximately 100 and 1000.  This makes the energetics of GRBs
consistent with the energetics of supernovae and merging compact
objects.

        Before \object{GRB~990123} GRB light curves were usually fit
by power laws of the form $f_{\nu}(t) = k t^{\alpha}$ where $k$ is a
normalization constant and $\alpha$ is between approximately $-1$ and
$-2$.  The steepest temporal decay was $\alpha = -2.5$ for the OA
associated with \object{GRB~971227} (Djorgovski et~al.~\cite{D98}).
The first evidence that the decay of the optical light curve from an
OA varies with time came from \object{GRB~990123}, where $\alpha$
rapidly steepened from $-1.10 \pm 0.03$ to $-1.65 \pm 0.06$ with the
break occurring $2.04 \pm 0.46$ days after the burst (Kulkarni
et~al.~\cite{KD99}).  Harrison et~al.~(\cite{HB99}), Israel
et~al.~(\cite{IMC1999}), and Stanek et~al.~(\cite{SGK1999}) found that
the optical light curve for \object{GRB~9900510} exhibited a similar
behavior, but with a slow transition between the two regimes.
Harrison et~al.~(\cite{HB99}) used a four-parameter fitting function
to find slopes of $-0.82 \pm 0.02$ before the break and $-2.18 \pm
0.05$ after the break with the break occurring $1.20 \pm 0.08$ days
after the burst.  However, Stanek et~al.~(\cite{SGK1999}) found slopes
of $-0.76 \pm 0.01$ before and $-2.40 \pm 0.02$ the break with the
break occurring $1.57 \pm 0.03$ days after the burst; and Israel
et~al.~(\cite{IMC1999}) found $-0.88 \pm 0.03$ before and $-2.68 \pm
0.13$ after with a break at $1.8 \pm 0.2$.  The latter two groups used
a different four-parameter fitting function from Harrison
et~al.~(\cite{HB99}), and the fits formally disagree with each other.
  
        In this paper we collect the published photometry for the OAs
associated with \object{GRB~990123} and \object{GRB~990510}, examine
three fitting functions for broken light curves, and use these fits to
constrain physical models for the burst.  We investigate the effects
of using different fitting functions on determining the time of the
break and the slope of the light curve before and after the break.
Our goal is to determine reliable light curve parameters for each GRB
and to use these parameters to distinguish between different physical
models for the nature of each OA\@.  Sect.~\ref{SECTION:fitting}
discusses the three fitting functions that we used to determine the
break time, and the slopes of each light curve.
Sect.~\ref{SECTION:grb990123_data} presents our fits to the light
curve of \object{GRB~990123}'s OA and an estimate of the slope of the
optical spectrum of the OA\@.  Sect.~\ref{SECTION:grb990510_data}
presents our fits to the light curve of \object{GRB~990510}'s OA and
an estimate of the slope of its optical spectrum.  We discuss the
effects of the choice of the fitting function on the determination of
the break time and slopes of the light curves for each burst.  In
Sect.~\ref{SECTION:models} we discuss various models for the decay of
the light curves, and in Sect.~\ref{SECTION:interpertation} we derive
some physical properties for the bursts based on our fits to the light
curves.  We have written these sections in such a way that they can be
used with data from other GRBs.

	We have adopted a cosmology with a Hubble parameter of
$H_0=65$ km s$^{-1}$ Mpc$^{-1}$, and assumed a density parameter of
$\Omega_0=0.2$ and no cosmological constant ($\Omega_\Lambda=0$)
throughout this paper.

 
\section{Fitting the Light Curves\label{SECTION:fitting}} 
 
        The choice of fitting function is arbitrary, but that choice
can affect the determination of the slopes and break time, and thus
can influence the interpretation of the light curve.  For example,
Jensen et~al.~(\cite{JFG2000}) find different break times and slopes
for the decay of the optical light curve of \object{GRB~000301C}
depending on whether a broken power law or the continuous function of
Stanek et~al.~(\cite{SGK1999}) is fit to the data.  We have chosen to
fit the light curves for the OAs associated with \object{GRB~990123}
and \object{GRB~990510} with three fitting functions.  The first of
these is a broken power law,
 
\begin{equation} 
\label{EQUATION:broken_power_law} 
f_{\nu}(t) = \left \{ 
        \begin{array}{lll} 
                f_{\nu}(t_b) {(t/t_b)}^{\alpha_1}, & 
                \mathrm{if}  &  t \le t_b \\ 
                f_{\nu}(t_b) {(t/t_b)}^{\alpha_2}, & 
                \mathrm{if}  &  t > t_b, 
        \end{array} 
             \right. 
\end{equation} 
 
\noindent 
where $f_{\nu}(t)$ is the flux in $\mu$Jy $t$ days after the burst,
and $t_b$ is the time that the light curve made the transition from a
slope of $\alpha_1$ to a slope of $\alpha_2$.  The slope goes as
$f_{\nu} \propto t^{\alpha_1}$ before the break and $f_{\nu} \propto
t^{\alpha_2}$ after the break with an instantaneous transition between
the two slopes.  This is the simplest form that can be fit to the
data, and it can be reduced to two linear functions (i.e.,
$\ln\left(f_{\nu}(t)\right) = \alpha_1 (t/t_b) +
\ln\left(f_{\nu}(t_b)\right)$ before the break and
$\ln\left(f_{\nu}(t)\right) = \alpha_2 (t/t_b) +
\ln\left(f_{\nu}(t_b)\right)$ after the break), which simplifies the
fitting process.
 
        The second fitting function is, 
 
\begin{equation} 
\label{EQUATION:harrison} 
f_\nu(t) = {f_\nu(t_b) \over 1 - e^{-1}} 
           {\left(\frac{t}{t_b}\right)}^{\alpha_1} 
           {{\left[1 - e^{-J}\right]} \over J}, 
\end{equation} 
 
\noindent 
where $J = {(t/t_b)}^{\alpha_1-\alpha_2}$.  This is the continuous
function used by Harrison et~al.~(\cite{HB99}).  It is differentiable
at $t = t_b$, and is generally more flexible than
Eq.~(\ref{EQUATION:broken_power_law}) is.  It is able to model the
slow transition between the $\alpha_1$ and $\alpha_2$ regimes, which
may be more physically realistic than the instantaneous transition
that is assumed by the broken power law.
 
        The third fitting function is, 
 
\begin{equation} 
f_\nu(t) = {2 f_\nu(t_b) {(t/t_b)}^{\alpha_2} 
           \over 
           1 + {(t/t_b)}^{\alpha_2 - \alpha_1}}, 
\label{EQUATION:stanek} 
\end{equation} 
 
\noindent 
which is the same as that used by Stanek et~al.~(\cite{SGK1999}) and
Israel et~al.~(\cite{IMC1999}), except that we have redefined
$\alpha_1$ and $\alpha_2$ to be consistent with our
Eqs.~(\ref{EQUATION:broken_power_law})~and~(\ref{EQUATION:harrison}).
This function has the same advantages over
Eq.~(\ref{EQUATION:broken_power_law}) that
Eq.~(\ref{EQUATION:harrison}) does.
 
        We wish to stress that none of these functional forms have
been derived from any physical principles, and that they are
\emph{not} based on any model for the GRB or its OA\@.  The three
forms were selected solely because they allow us to parameterize the
observed light curves for \object{GRB~990123} and \object{GRB~990510}.
We used CERN's {\sc Minuit} function minimization package, and a
chi-square minimization scheme, to simultaneously solve for the four
free parameters ($\alpha_1$, $\alpha_2$, $t_b$, and $f_{\nu}(t_b)$)
and to compute the formal 1$\sigma$ errors in the fit for each
parameter.

\section{GRB 990123 Data\label{SECTION:grb990123_data}} 
 
\subsection{The Photometry\label{SECTION:grb990123_photometry}} 
 
        We collected the published optical observations of the OA
associated with \object{GRB~990123} from six sources: Castro-Tirado
et~al.~(\cite{CT99}), Fruchter et~al.~(\cite{FT99a}), Galama
et~al.~(\cite{GB99}), Holland \& Hjorth~(\cite{HH99}), Kulkarni
et~al.~(\cite{KD99}), and Fruchter et~al.~(\cite{FT99b}).  These data
consists of 88 observations in the Johnson $B$-, and $V$-bands,
Kron-Cousins $R$-, and $I$-bands, and Gunn $r$-band.  The observations
were taken at 19 telescopes over a two month period after the burst.
In cases where the same observation was reported in two or more
sources, and the quoted magnitudes disagreed with each other, we
discarded both observations.  This was done because we had no reason
to favour one source over the other.  In cases where a Gunn $r$-band
magnitude was derived from an $R$-band magnitude we discarded the
derived Gunn $r$-band data and kept the original $R$-band data.
 
        We converted the published magnitudes to fluxes using the
photometric zero points, $f_{\nu,0}$ from Fukugita
et~al.~(\cite{FS95}).  These zero points, and the central wavelengths
for each band pass, are listed in Table~\ref{TABLE:filters}.  In order
to convert the published photometry to a uniform set of fluxes we
applied our own corrections for the light of the probable host galaxy
(Holland \& Hjorth~\cite{HH99}).  In cases where the light from the
galaxy had already been subtracted in the published photometry we
added it back using the magnitude for the galaxy that was given in the
source then subtracted a uniform estimate of the light from the galaxy
based on the data of Holland \& Hjorth~(\cite{HH99}) ($V_0 = 24.25 \pm
0.07$, see Table~\ref{TABLE:mag_gal}).  Castro-Tirado
et~al.~(\cite{CT99}) used $U\!BV\!RI$ photometry to determine a
spectral energy distribution of $f_{\nu} \propto
\nu^{\beta_\mathrm{gal}} = \mathrm{constant}$, i.e.,
$\beta_\mathrm{gal} \approx 0$, for the host galaxy.  Therefore, we
converted the Holland \& Hjorth~(\cite{HH99}) $V$-band flux for the
galaxy to the other band passes using $\mathrm{mag} = V +
K_\mathrm{mag} - K_V - 2.5\beta\log_{10}(\nu_\mathrm{mag}/\nu_V)$ and
$\beta_\mathrm{gal} = 0$ where $K_V$ and $K_\mathrm{mag}$ are the
photometric zero points of the two band passes.  The Castro-Tirado
et~al.~(\cite{CT99}) photometry, and our adopted (dereddened)
magnitudes and fluxes for the probable host galaxy, are listed in
Table~\ref{TABLE:mag_gal}.  A flat spectrum is consistent with the
probable host galaxy being a starburst galaxy.
 
\begin{table} 
\begin{center} 
\caption{The central wavelengths and photometric zero points for each 
band pass (from Fukugita et~al.~\cite{FS95}).} 
\smallskip 
\begin{tabular}{ccc} 
\hline 
\hline 
Filter & $\lambda_0$ ({\AA}) & $f_{\nu,0}$ (erg cm$^{-1}$ s$^{-1}$ Hz$^{-1}$) \\ 
\hline 
   $B$   &  4448 & $4.02 \times 10^{-20}$ \\ 
   $V$   &  5505 & $3.59 \times 10^{-20}$ \\ 
Gunn $r$ &  6538 & $2.96 \times 10^{-20}$ \\ 
   $R$   &  6588 & $3.02 \times 10^{-20}$ \\ 
   $I$   &  8060 & $2.38 \times 10^{-20}$ \\ 
\hline 
\hline 
\end{tabular} 
\label{TABLE:filters} 
\end{center} 
\end{table} 
 
\begin{table} 
\begin{center} 
\caption{The observed magnitudes (from Castro-Tirado 
et~al.~\cite{CT99}), and the adopted magnitudes and fluxes for the 
probable host galaxy of \object{GRB~990123}.} 
\smallskip 
\begin{tabular}{cccc} 
\hline 
\hline 
Filter & Castro-Tirado & adopted & adopted \\ 
       &      mag      & mag     & flux ($\mu$Jy) \\ 
\hline 
   $B$   & $24.23 \pm 0.10$ & $24.35 \pm 0.11$ & $0.729 \pm 0.074$ \\ 
   $V$   & $24.20 \pm 0.15$ & $24.25 \pm 0.07$ & $0.716 \pm 0.046$ \\ 
Gunn $r$ &      $\cdots$    & $24.05 \pm 0.07$ & $0.711 \pm 0.046$ \\ 
   $R$   & $23.77 \pm 0.10$ & $24.07 \pm 0.08$ & $0.709 \pm 0.052$ \\ 
   $I$   & $23.65 \pm 0.16$ & $23.83 \pm 0.08$ & $0.701 \pm 0.052$ \\ 
\hline 
\hline 
\end{tabular} 
\label{TABLE:mag_gal} 
\end{center} 
\end{table} 
 
        The Galactic reddening in the direction of \object{GRB~990123}
is $E_{B\!-\!V} = 0.016 \pm 0.020$ (Schlegel et~al.~\cite{SF98}).  We
assumed that there is no extinction in the host galaxy, but this is
probably not a good assumption since it is unlikely that a starburst
galaxy is devoid of dust.  Reddening due to dust in the host galaxy
may result in us underestimating the flux from the host galaxy, and
thus overestimating the flux from the OA\@.  To avoid this uncertainty
we have fit light curves to each band pass separately since the
systematic error in the flux introduced by the unknown extinction will
only affect the zero-point of the flux in each bandpass, not the
slopes of the light curves.
 
        The fits for each band pass, and each fitting function, are
given in Table~\ref{TABLE:grb990123_fits}.  The uncertainties in the
parameters are the formal 1$\sigma$ errors in the fit and do not
include contributions from the covariance between the parameters.  The
slopes are not strongly correlated with each other, but they are
moderately correlated with the location of the break.  None of the
residuals show any time dependence.
 
\begin{table*} 
\begin{center} 
\caption{The parameters of the best-fitting light curves for the
photometry of the OA associated with \object{GRB~990123} in each band
pass.  Eq.~(\ref{EQUATION:broken_power_law}) is a broken power law,
Eq.~(\ref{EQUATION:harrison}) is the continuous function of Harrison
et~al.~(\cite{HB99}), and Eq.~(\ref{EQUATION:stanek}) is the
continuous function of Stanek et~al.~(\cite{SGK1999}).}
\smallskip 
\begin{tabular}{cccccccc} 
\hline 
\hline 
Eq. & Filter & $\alpha_1$ & $\alpha_2$ & $t_b$ (days) & $f_{\nu}(t_b)$ ($\mu$Jy) & $\chi^2_{\mathrm{DOF}}$ & DOF \\ 
\hline 
({\ref{EQUATION:broken_power_law}})
&      $B$ & $-1.11 \pm 0.12$ & $-1.81 \pm 0.14$ & $1.57 \pm 0.51$ &  $9.32 \pm   4.38$ & 1.0224 &  6 \\ 
&      $V$ & $-1.28 \pm 0.48$ & $-1.58 \pm 0.16$ & $1.71 \pm 2.20$ &  $9.09 \pm  17.72$ & 0.7251 &  8 \\ 
& Gunn $r$ & $-1.11 \pm 0.12$ & $-2.01 \pm 0.16$ & $2.11 \pm 0.83$ &  $6.22 \pm   3.94$ & 2.6654 &  4 \\ 
&      $R$ & $-1.17 \pm 0.30$ & $-1.57 \pm 0.11$ & $1.70 \pm 0.22$ & $10.74 \pm   3.01$ & 0.9111 & 31 \\ 
&      $I$ & $-1.10 \pm 0.42$ & $-1.64 \pm 0.10$ & $1.36 \pm 0.70$ & $15.72 \pm  10.83$ & 0.6922 &  1 \\ 
\hline 
({\ref{EQUATION:harrison}})
&      $B$ & $-0.83 \pm 0.38$ & $-1.81 \pm 0.25$ & $0.76 \pm 0.93$ & $21.89 \pm  34.83$ & 0.5445 &  6 \\ 
&      $V$ & $-0.76 \pm 1.82$ & $-1.57 \pm 0.19$ & $0.39 \pm 0.64$ & $55.90 \pm 131.49$ & 0.8296 &  8 \\ 
& Gunn $r$ & $-0.91 \pm 0.24$ & $-2.11 \pm 0.31$ & $1.45 \pm 1.34$ &  $9.34 \pm  12.57$ & 3.6172 &  4 \\ 
&      $R$ & $-0.73 \pm 1.16$ & $-1.57 \pm 0.22$ & $0.54 \pm 0.84$ & $40.55 \pm  71.25$ & 0.8444 & 31 \\ 
&      $I$ & $19.07 \pm\cdots$& $-1.61 \pm 0.06$ & $0.70 \pm 0.80$ & $26.00 \pm  47.72$ & 0.7975 &  1 \\ 
\hline 
({\ref{EQUATION:stanek}})
&      $B$ & $-0.75 \pm 0.70$ & $-1.90 \pm 0.43$ & $0.92 \pm 2.11$ & $17.22 \pm  53.45$ & 0.5647 &  6 \\ 
&      $V$ & $-1.12 \pm 0.60$ & $-1.71 \pm 0.26$ & $1.40 \pm 4.29$ & $11.09 \pm  48.05$ & 0.8678 &  8 \\ 
& Gunn $r$ & $-0.87 \pm 0.38$ & $-2.25 \pm 0.47$ & $1.78 \pm 2.46$ &  $6.78 \pm  14.37$ & 3.9109 &  4 \\ 
&      $R$ & $-0.56 \pm 3.23$ & $-1.63 \pm 0.68$ & $0.54 \pm 2.63$ & $40.64 \pm 214.63$ & 0.8722 & 31 \\ 
&      $I$ &  $2.55 \pm 9.37$ & $-1.62 \pm 0.10$ & $0.54 \pm 0.24$ & $36.76 \pm  30.45$ & 0.7451 &  1 \\ 
\hline 
\hline 
\end{tabular} 
\label{TABLE:grb990123_fits} 
\end{center} 
\end{table*} 
 
\begin{figure} 
\resizebox{\hsize}{!}{\includegraphics{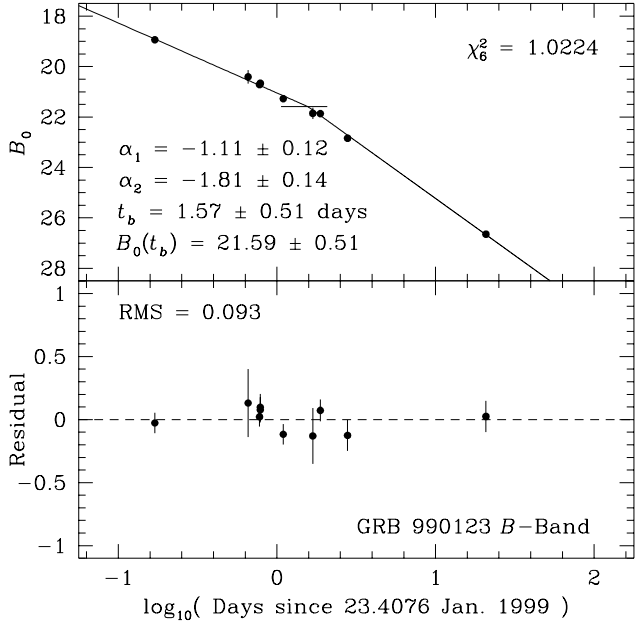}} 
\caption{The upper panel shows the best-fitting broken power law
(Eq.~(\ref{EQUATION:broken_power_law})) for the \object{GRB~990123}
$B$-band photometry.  The horizontal bar shows the $1 \sigma$
uncertainty in the time of the break.  The lower panel shows the
residuals in the fit.  The residuals are defined as $(B_\mathrm{fit} -
B_\mathrm{obs})$.  The uncertainties in the residuals are the
uncertainties in the observed data.  The magnitudes have been
corrected for Galactic extinction in the direction of
\object{GRB~990123}, but not for extinction in the host galaxy.}
\label{FIGURE:grb990123_B} 
\end{figure}

\begin{figure} 
\resizebox{\hsize}{!}{\includegraphics{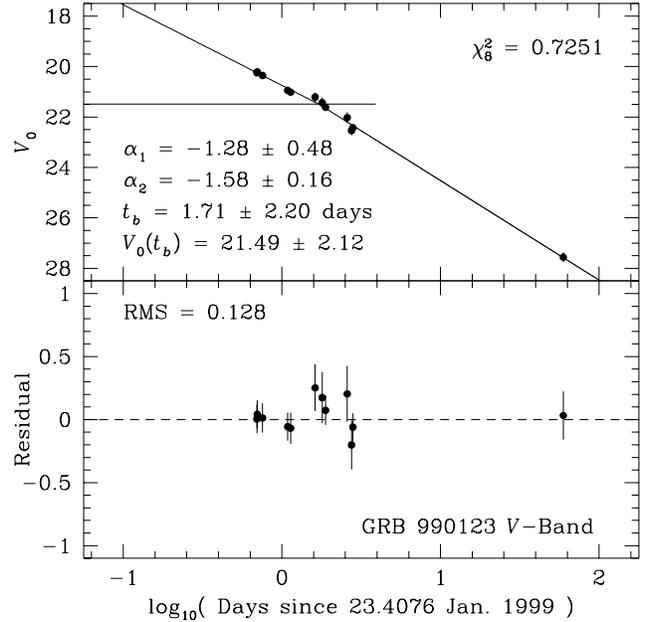}} 
\caption{This figure shows the best-fitting broken power law, and the 
residuals, for the \object{GRB~990123} $V$-band data.  The large 
uncertainty in the time of the break ($t_b = 1.71 \pm 2.20$ days) 
indicates that the data could be well fit by a single power law. } 
\label{FIGURE:grb990123_V} 
\end{figure}

\begin{figure} 
\resizebox{\hsize}{!}{\includegraphics{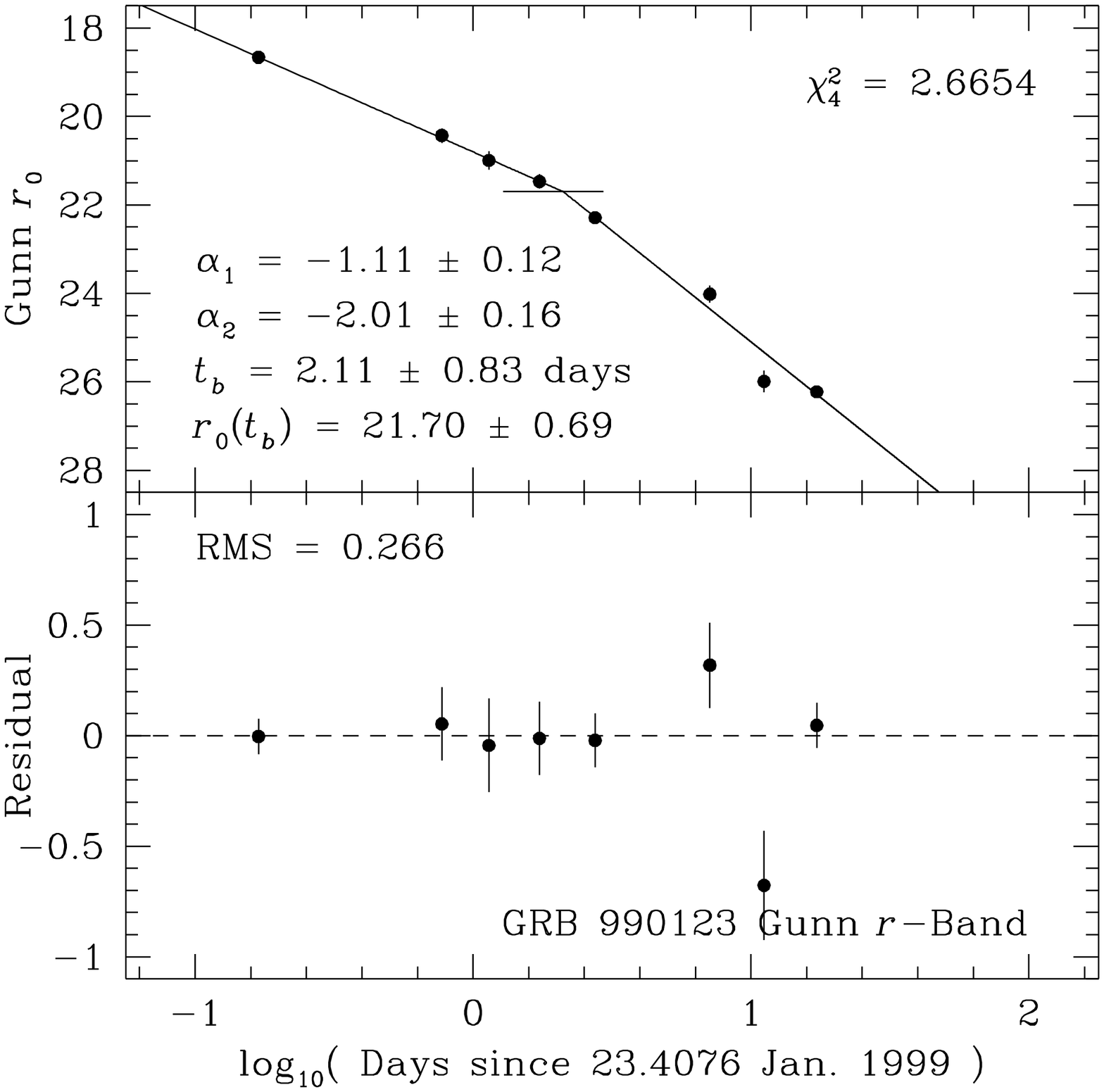}} 
\caption{This figure shows the best-fitting broken power law, and the 
residuals, for the \object{GRB~990123} Gunn $r$-band data.} 
\label{FIGURE:grb990123_r} 
\end{figure}

\begin{figure} 
\resizebox{\hsize}{!}{\includegraphics{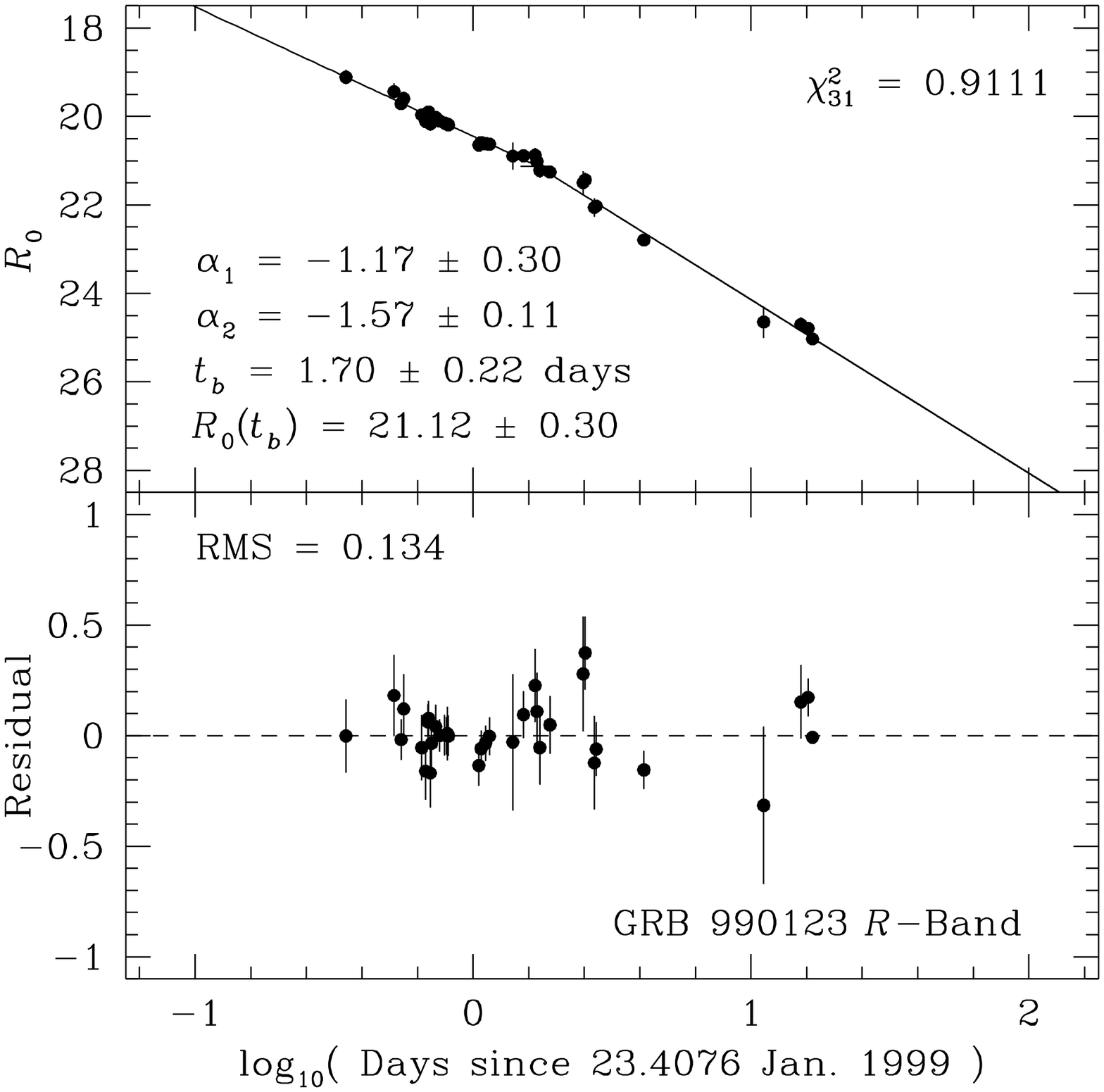}} 
\caption{This figure shows the best-fitting broken power law, and the 
residuals, for the \object{GRB~990123} $R$-band data.} 
\label{FIGURE:grb990123_R} 
\end{figure}

\begin{figure} 
\resizebox{\hsize}{!}{\includegraphics{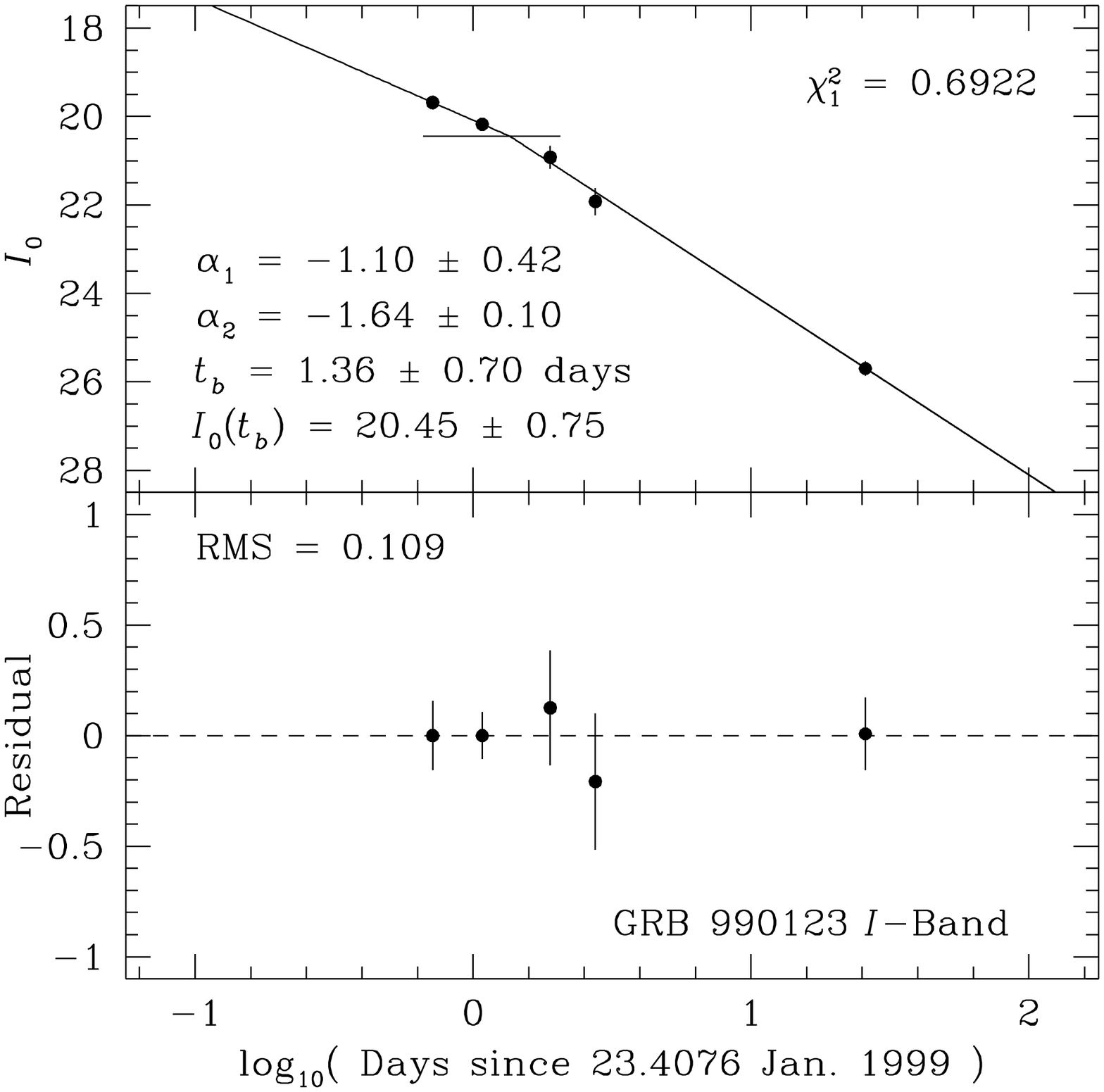}} 
\caption{This figure shows the best-fitting broken power law, and the 
residuals, for the \object{GRB~990123} $I$-band data.} 
\label{FIGURE:grb990123_I} 
\end{figure}

        The weighted mean values for each parameter, their standard
errors (SE), and the $\chi^2$ per degree of freedom
($\chi^2_\mathrm{DOF}$) values, are listed in
Table~\ref{TABLE:grb990123_mean}, for each fitting function.  The
$BVrRI$ data was used to compute the mean values of the fits to the
broken power-law (Eq.~(\ref{EQUATION:broken_power_law})) and the
$BVrR$ data was used to compute the mean values of the fits to the two
continuous functions
(Eqs.~(\ref{EQUATION:harrison})~and~(\ref{EQUATION:stanek})).  The
$I$-band data were not used when computing the latter two sets of mean
values because the fits to the $I$-band data were not reliable for $t
< t_b$.  The mean values of $\alpha_1$ and $\alpha_2$ agree to within
one $\sigma$ regardless of which function is fit to the data, but the
mean value of $t_b$ is significantly smaller for
Eq.~(\ref{EQUATION:harrison}) (Harrison's function) than it is for the
other two fitting functions.  However, the broken power law
(Eq.~(\ref{EQUATION:broken_power_law})) gives considerably less
scatter in the values of each parameter, particularly $t_b$, for
different band passes than the other two fitting functions do.
Therefore, we conclude that the broken power law gives a better fit to
the \object{GRB~990123} light curves than either of the continuous
functions do.  This suggests that the break in the light curve may
have been very rapid.  The 1$\sigma$ error in the break time for a
broken power law is 0.19 days (see Table~\ref{TABLE:grb990123_mean}),
which suggests that the break may have occurred over a period of only
a few hours.  Figs.~\ref{FIGURE:grb990123_B}--\ref{FIGURE:grb990123_I}
show the data for each band pass with the best-fitting broken power
laws superimposed.
 
\begin{table} 
\begin{center} 
\caption{The mean values of $\alpha_1$, $\alpha_2$, and $t_b$ for the
best fits of Eq.~(\ref{EQUATION:broken_power_law}) (a broken power
law) to the $BVrRI$ data, and of Eq.~(\ref{EQUATION:harrison})
(Harrison's function) and Eq.~(\ref{EQUATION:stanek}) (Stanek's
function) to the $BVrR$ data, for the OA associated with
\object{GRB~990123}.  $P$ is the probability that each parameter
depends on wavelength (see the text).}
\smallskip 
\begin{tabular}{ccccc} 
\hline 
\hline 
Eq. & Parameter & Mean $\pm$ SE & $\chi^2_4$ & $P$ \\ 
\hline 
({\ref{EQUATION:broken_power_law}})
& $\alpha_1$   & $-1.12 \pm 0.08$ & 0.0387 & 0.0028 \\ 
& $\alpha_2$   & $-1.69 \pm 0.06$ & 1.6616 & 0.8442 \\ 
& $t_b$ (days) &  $1.68 \pm 0.19$ & 0.1331 & 0.0297 \\ 
\hline 
({\ref{EQUATION:harrison}})
& $\alpha_1$   & $-0.88 \pm 0.20$ & 0.0180 & 0.0033 \\ 
& $\alpha_2$   & $-1.70 \pm 0.12$ & 0.9197 & 0.5697 \\ 
& $t_b$ (days) &  $0.61 \pm 0.42$ & 0.1814 & 0.0909 \\  
\hline 
({\ref{EQUATION:stanek}})
& $\alpha_1$   & $-0.91 \pm 0.29$ & 0.0658 & 0.0220 \\ 
& $\alpha_2$   & $-1.83 \pm 0.19$ & 0.3748 & 0.2288 \\ 
& $t_b$ (days) &  $1.11 \pm 1.30$ & 0.0446 & 0.0125 \\ 
\hline 
\hline 
\end{tabular} 
\label{TABLE:grb990123_mean} 
\end{center} 
\end{table} 
 
        To test if the break time depends on the wavelength we
computed the probabilities that the break times in each band pass are
consistent with the mean break time.  If the individual break times
are not consistent with the mean then this would suggest that the
break between the $\alpha_1$ and $\alpha_2$ regimes depends on
wavelength, which would be inconsistent with the predictions of the
relativistic fireball model.  This method makes no assumptions about
the physical origin of the light curve.  We checked if the break time
was dependent on wavelength by evaluating the $\chi^2$ statistic for
the weighted mean value of $t_b$.  The $\chi^2_\mathrm{DOF}$ value,
and $P$, the probability that we can reject the null hypothesis that
the break time seen in each band pass is consistent with the mean
break time (i.e.\ $P$ is the probability that the break time changes
with wavelength), are listed in Table~\ref{TABLE:grb990123_mean}.
Similar calculations were done for the slopes $\alpha_1$ and
$\alpha_2$.  The $P$ values are all less than 0.9, so we conclude that
there is no evidence that the break time, or the slopes, vary with
wavelength.

\subsection{The Spectrum of the Optical Transient\label{SECTION:grb990123_beta_OA}} 
 
        We have assumed that the OA for \object{GRB~990123} has a 
power-law spectrum of the form 
 
\begin{equation} 
f_{\nu}(\nu) = f_{\nu}(10^{15}\,\mathrm{Hz}) 
     {\left({\nu \over 10^{15}\,\mathrm{Hz}}\right)}^{\beta_\mathrm{OA}}, 
\label{EQUATION:spectrum_OA} 
\end{equation} 
 
\noindent 
where $f_{\nu}(\nu)$ is the flux, in $\mu$Jy, at the frequency $\nu$,
and $\beta_\mathrm{OA}$ is the spectral index.  The choice of
$10^{15}$ Hz as a reference frequency is arbitrary, but this value
gives $\nu/10^{15} \approx 1$, which results in stable fits of the
flux data to Eq.~(\ref{EQUATION:spectrum_OA}).
 
\begin{figure} 
\resizebox{\hsize}{!}{\includegraphics{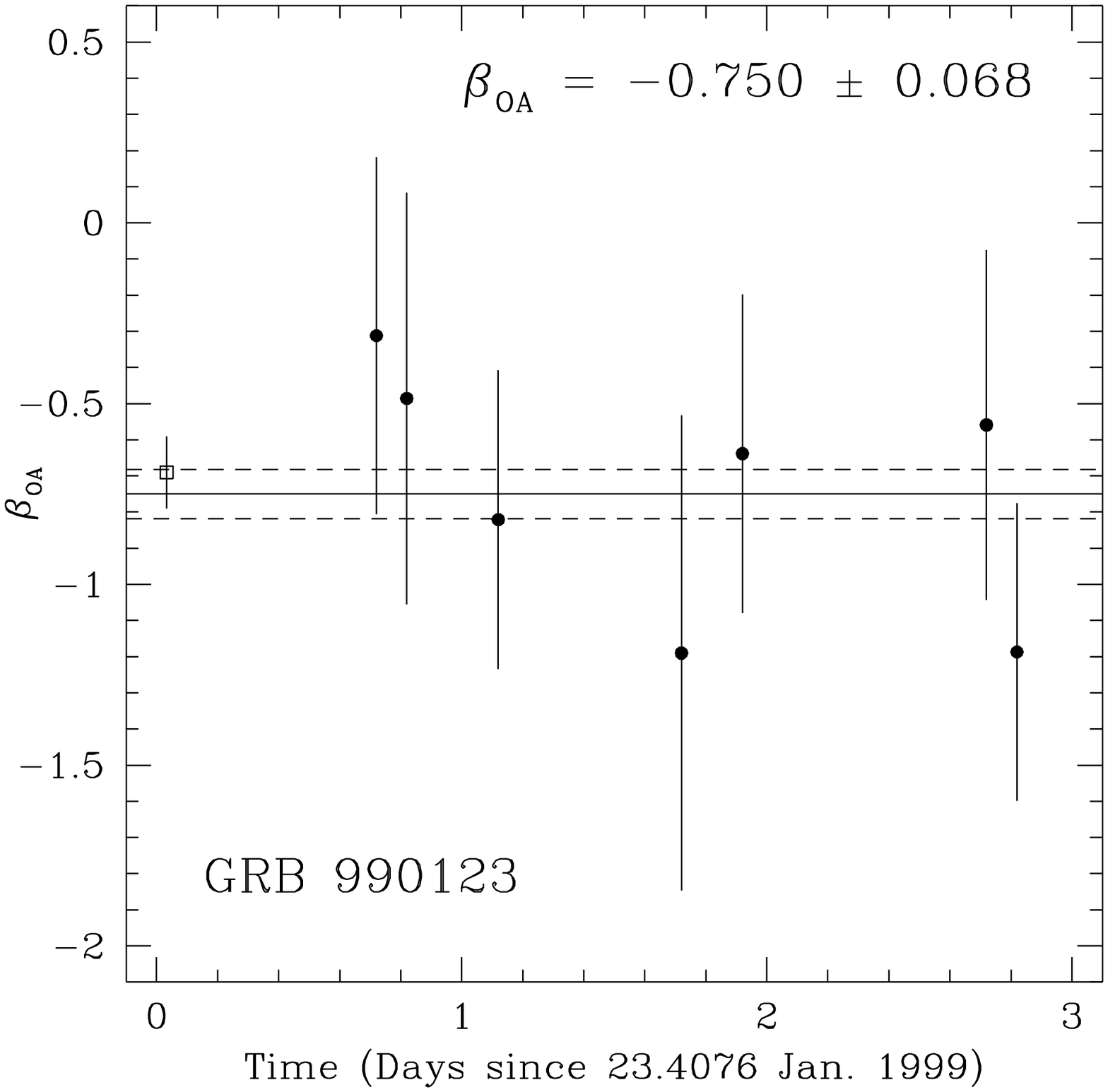}} 
\caption{This figure shows $\beta_\mathrm{OA}$, the spectral index of
the light from the OA associated with \object{GRB~990123} (filled
circles), as a function of time.  The error bars are defined as
$\sigma_\beta = \sqrt{\sum_{i=1}^N {(\sigma_{f,i}/f_i)}^2}$ where
$\sigma_{f,i}$ is the uncertainty in each flux value, $f_i$.  The
solid line shows the weighted mean value of $\beta_\mathrm{OA}$ while
the dashed lines show the $1 \sigma$ uncertainty in
$\overline{\beta_\mathrm{OA}}$.  The open square shows the Andersen
et~al.~(\cite{AC99}) spectral value.  The data are consistent with a
constant $\beta_\mathrm{OA}$ of $-0.750 \pm 0.068$.  There is no
evidence that $\beta_\mathrm{OA}$ varies with time.}
\label{FIGURE:grb990123_beta} 
\end{figure}

        We estimated the spectrum at a series of times by taking the
$B$-, $V$-, $R$-, and $I$-band data in 0.1-day intervals and fitting
it to Eq.~(\ref{EQUATION:spectrum_OA}).  Only intervals that contained
at least three observations spanning two or more filters were
considered.  Fig.~\ref{FIGURE:grb990123_beta} shows the values of
$\beta_\mathrm{OA}$ determined this way.  The weighted mean spectral
index is $\overline{\beta_\mathrm{OA}} = -0.750 \pm 0.068$ (SE) with
$\chi^2_7 = 0.4711$.  The low $\chi^2_\mathrm{DOF}$ value suggests
that we can reject the null hypothesis, that the spectral index is
constant with time, at the 17\% confidence level.  This is too low to
be able to reject the null hypothesis, so we conclude that there is no
evidence for a variable $\beta_\mathrm{OA}$.  Our value of
$\beta_\mathrm{OA}$ is consistent with the direct measurement of
Andersen et~al.~(\cite{AC99}).  They found $\beta_\mathrm{OA} = -0.69
\pm 0.10$ between 4000 {\AA} $\le \lambda \le$ 5700 {\AA} from a
spectrum of the OA that was taken 0.8 hours after the burst.
 
\section{GRB 990510 Data\label{SECTION:grb990510_data}} 
 
\subsection{The Photometry\label{SECTION:grb990510_photometry}} 
 
        We collected the published optical observations of the OA
associated with \object{GRB~990510} from eleven sources: Beuermann
et~al.~(\cite{BH99}), Covino et~al.~(\cite{CL99}), Fruchter
et~al.~(\cite{FF99}), Galama et~al.~(\cite{GVR1999}), Harrison
et~al.~(\cite{HB99}), Marconi et~al.~(\cite{MI99a},~\cite{MI99b}),
Pietrzy{\'n}ski \& Udalski~(\cite{PU99a},~\cite{PU99b},~\cite{PU99c}),
and Stanek et~al.~(\cite{SGK1999}).  These data consists of 182
observations in the Johnson $V$-band, and in the Kron-Cousins $R$-,
and $I$-bands.  The observations were taken at seven telescopes up to
40 days after the burst.  The published magnitudes were converted to
fluxes in the same manner as described in
Sect.~\ref{SECTION:grb990123_photometry}.  No corrections were made
for a host galaxy since no host galaxy has been observed for this
burst to a limiting magnitude of $R = 27.6$ (Beuermann
et~al.~\cite{BH99})\footnote{Recently Bloom~(\cite{B2000}) and
Fruchter et~al~(\cite{FHP2000}) reported evidence for a faint ($V =
28.5 \pm 0.5$) host galaxy.  This corresponds to a flux of $\approx
0.014$ $\mu$Jy in the V band, which is not sufficient to affect the
results of our fits.}.  The Galactic reddening in the direction of
\object{GRB~990510} is $E_{B\!-\!V} = 0.203 \pm 0.020$ (Schlegel
et~al.~\cite{SF98}).  We assumed that the extinction in the host
galaxy was $A_V = 0$.
 
        The fits for each band pass, and each fitting function, are
given in Table~\ref{TABLE:grb990510_fits}.  The uncertainties in the
parameters are the formal 1$\sigma$ errors in the fits and do not
include contributions from the covariance between the parameters.  The
residuals for all three fitting functions show some structure near the
location of the break.  This suggests that none of the three fitting
functions which we used in this paper accurately reproduce the
temporal behavior of the light curve during the transition from the
$\alpha_1$ regime to the $\alpha_2$ regime.  The last two $V$-band
observations, and the last $R$-band observation, were derived from
images taken with the \emph{HST}/STIS and were transformed from the
50CCD (clear) aperture to the $V$ and $R$ bands assuming a spectral
index of $\beta_\mathrm{OA} = -0.6$ (Fruchter et~al.~\cite{FF99}).
These observations were taken approximately one month after the other
observations, so the large residuals for these three data points may
be due to a change in the shape of the OA's spectrum during that
period, or systematic uncertainties in transforming the 50CCD
magnitudes to $VR$ magnitudes.
 
\begin{table*} 
\begin{center} 
\caption{The parameters of the best-fitting light curves for the
photometry of the OA associated with \object{GRB~990510} in each band
pass.  Eq.~(\ref{EQUATION:broken_power_law}) is a broken power law,
Eq.~(\ref{EQUATION:harrison}) is the continuous function of Harrison
et~al.~(\cite{HB99}), and Eq.~(\ref{EQUATION:stanek}) is the continuous
function of Stanek et~al.~(\cite{SGK1999}).}
\smallskip 
\begin{tabular}{cccccccc} 
\hline 
\hline 
Eq. & Filter & $\alpha_1$ & $\alpha_2$ & $t_b$ & $f_{\nu}(t_b)$ & $\chi^2_{\mathrm{DOF}}$ & DOF \\ 
\hline 
({\ref{EQUATION:broken_power_law}})
& $V$ & $-0.78 \pm 0.12$ & $-2.08 \pm 0.20$ & $1.23 \pm 0.22$ &  $62.72 \pm  13.20$ & 1.5183 & 63 \\ 
& $R$ & $-0.69 \pm 0.11$ & $-1.72 \pm 0.13$ & $0.87 \pm 0.12$ &  $96.06 \pm  15.50$ & 1.2569 & 57 \\ 
& $I$ & $-1.16 \pm 0.34$ & $-1.73 \pm 0.29$ & $1.33 \pm 0.66$ &  $58.67 \pm  41.41$ & 0.9355 & 50 \\ 
\hline 
({\ref{EQUATION:harrison}})
& $V$ & $-0.62 \pm 0.19$ & $-2.16 \pm 0.29$ & $1.05 \pm 0.47$ &  $65.54 \pm  37.73$ & 0.6395 & 63 \\ 
& $R$ & $-0.46 \pm 0.20$ & $-1.85 \pm 0.26$ & $0.70 \pm 0.35$ & $114.54 \pm  63.24$ & 0.7292 & 57 \\ 
& $I$ & $-0.67 \pm 1.79$ & $-1.84 \pm 1.22$ & $0.80 \pm 2.26$ & $106.44 \pm 351.17$ & 0.8679 & 50 \\ 
\hline 
({\ref{EQUATION:stanek}})
& $V$ & $-0.58 \pm 0.23$ & $-2.32 \pm 0.37$ & $1.25 \pm 0.71$ &  $50.70 \pm  39.84$ & 0.4218 & 63 \\ 
& $R$ & $-0.40 \pm 0.29$ & $-2.00 \pm 0.36$ & $0.81 \pm 0.60$ &  $96.02 \pm  84.48$ & 0.9763 & 57 \\ 
& $I$ & $-0.51 \pm 2.08$ & $-2.02 \pm 1.16$ & $0.98 \pm 2.85$ &  $83.55 \pm 310.43$ & 0.8547 & 50 \\ 
\hline 
\hline 
\end{tabular} 
\label{TABLE:grb990510_fits} 
\end{center} 
\end{table*} 
 
\begin{figure} 
\resizebox{\hsize}{!}{\includegraphics{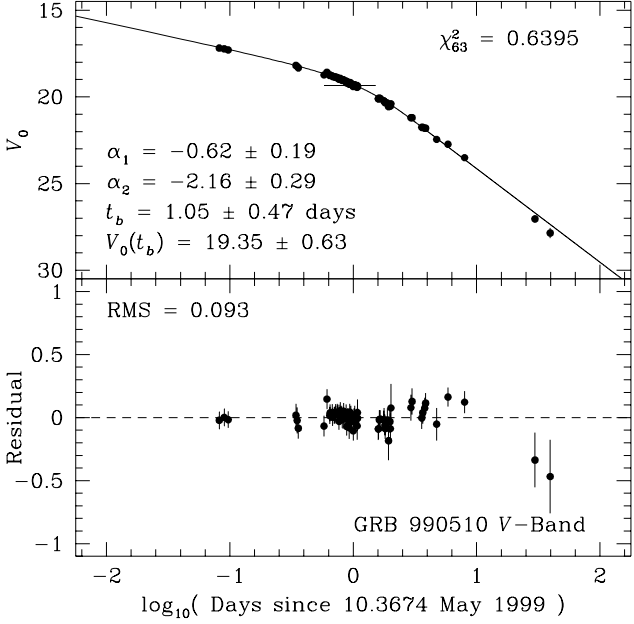}} 
\caption{The upper panel shows the best-fitting continuous function
(Eq.~(\ref{EQUATION:harrison}), Harrison's function) for the
\object{GRB~990510} $V$-band photometry.  The horizontal bar shows the
$1 \sigma$ uncertainty in the time of the break.  The lower panel
shows the residuals in the fit in the sense $(V_\mathrm{fit} -
V_\mathrm{obs})$.  The magnitudes have been corrected for Galactic
extinction in the direction of \object{GRB~990510}.}
\label{FIGURE:grb990510_V} 
\end{figure}

\begin{figure} 
\resizebox{\hsize}{!}{\includegraphics{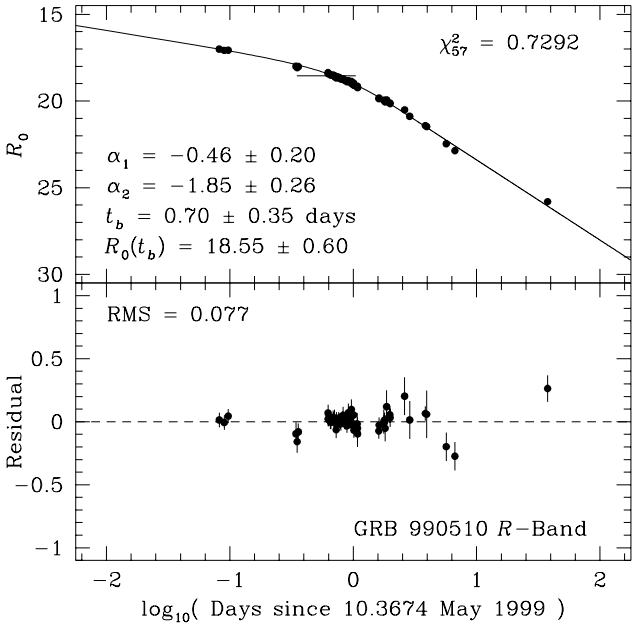}} 
\caption{This figure shows the best-fitting continuous function
(Eq.~(\ref{EQUATION:harrison})), and the residuals, for the
\object{GRB~990510} $R$-band data.}
\label{FIGURE:grb990510_R} 
\end{figure}

\begin{figure} 
\resizebox{\hsize}{!}{\includegraphics{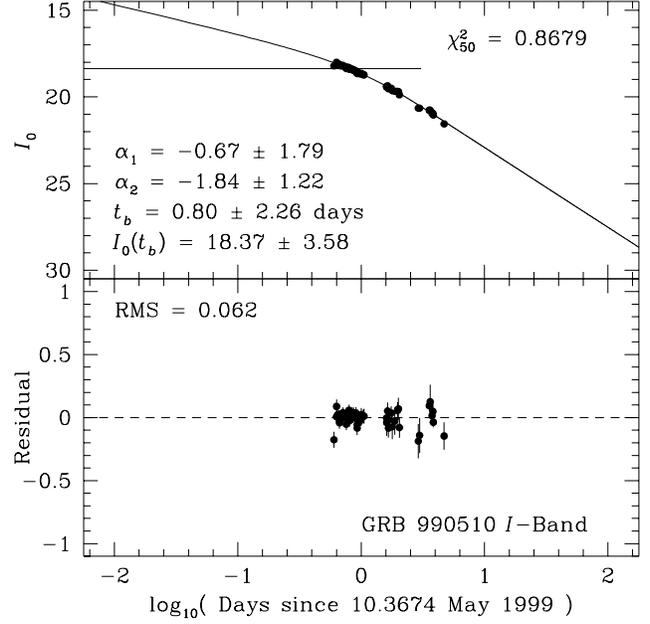}} 
\caption{This figure shows the best-fitting continuous function
(Eq.~(\ref{EQUATION:harrison})), and the residuals, for the
\object{GRB~990510} $I$-band data.  The large uncertainty in the time
of the break ($t_b = 0.80 \pm 2.26$ days) indicates that the data can
be well fit by a single power law.}
\label{FIGURE:grb990510_I} 
\end{figure}

        The weighted mean values for each parameter, their standard
errors, and the $\chi^2_\mathrm{DOF}$ values are listed in
Table~\ref{TABLE:grb990510_mean} for the fits to each fitting
function.  The mean values of $\alpha_1$, $\alpha_2$, and $t_b$ agree
to within approximately $1.5 \sigma$ regardless of which function is
fit to the data.  However, there are systematic variations in the
residuals of the fits of Eq.~(\ref{EQUATION:broken_power_law}) (the
broken power law) to the $V$- and $R$-band data.  This indicates that
one of the continuous functions
(Eq.~(\ref{EQUATION:harrison})~or~(\ref{EQUATION:stanek})) provides a
better fit to the data than the broken power law
(Eq.~(\ref{EQUATION:broken_power_law})) does, which in turn suggests
that the break in the light curve was not nearly instantaneous as was
the case with \object{GRB~990123}, but may have occurred over a period
of approximately one day.  Eq.~(\ref{EQUATION:harrison}) (Harrison's
function) gives the smallest residuals, and the least scatter in the
values of the parameters in different band passes, so we have adopted
it as the best fitting function.  The $P$ values in
Table~\ref{TABLE:grb990510_mean} are all less than 0.7, so we conclude
that there is no evidence for any variation in $\alpha_1$, $\alpha_2$,
or $t_b$ with wavelength.
Figs.~\ref{FIGURE:grb990510_V}--\ref{FIGURE:grb990510_I} show the data
for each band pass with the best-fitting continuous functions
superimposed.
 
\begin{table} 
\begin{center} 
\caption{The mean values of $\alpha_1$, $\alpha_2$, and $t_b$ for the
best fits of Eq.~(\ref{EQUATION:broken_power_law}) (a broken power
law), Eq.~(\ref{EQUATION:harrison}) (Harrison's function), and
Eq.~(\ref{EQUATION:stanek}) (Stanek's function) to the $V\!RI$ data
for the OA associated with \object{GRB~990510}.  $P$ is the
probability that each parameter depends on wavelength (see the text).}
\smallskip 
\begin{tabular}{ccccc} 
\hline 
\hline 
Eq. & Parameter & Mean $\pm$ SE & $\chi^2_3$ & $P$ \\ 
\hline 
({\ref{EQUATION:broken_power_law}})
& $\alpha_1$ & $-0.75 \pm 0.18$ & 1.8114 & 0.5957 \\ 
& $\alpha_2$ & $-1.81 \pm 0.10$ & 1.1878 & 0.6951 \\ 
& $t_b$      &  $0.96 \pm 0.10$ & 1.1913 & 0.6962 \\ 
\hline 
({\ref{EQUATION:harrison}})
& $\alpha_1$ & $-0.54 \pm 0.14$ & 0.1707 & 0.1569 \\ 
& $\alpha_2$ & $-1.98 \pm 0.19$ & 0.3239 & 0.2767 \\ 
& $t_b$     &  $0.82 \pm 0.28$ & 0.1784 & 0.1634 \\ 
\hline 
({\ref{EQUATION:stanek}})
& $\alpha_1$ & $-0.51 \pm 0.18$ & 0.1182 & 0.1115 \\ 
& $\alpha_2$ & $-2.15 \pm 0.25$ & 0.1986 & 0.1801 \\ 
& $t_b$      &  $0.99 \pm 0.45$ & 0.1120 & 0.1060 \\ 
\hline 
\hline 
\end{tabular} 
\label{TABLE:grb990510_mean} 
\end{center} 
\end{table} 
 
\subsection{The Spectrum of the Optical Transient\label{SECTION:grb990510_beta_OA}} 
 
\begin{figure} 
\resizebox{\hsize}{!}{\includegraphics{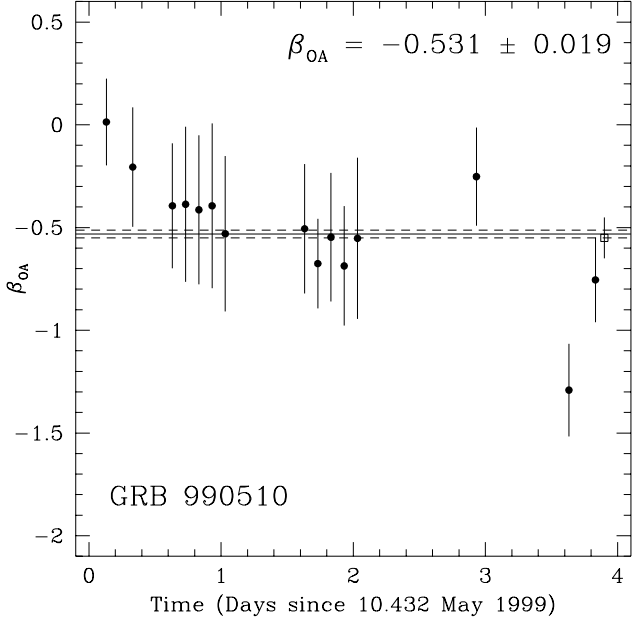}} 
\caption{This figure shows $\beta_\mathrm{OA}$, the spectral index of
the light from the OA associated with \object{GRB~990510} (filled
circles), as a function of time.  The solid line shows the weighted
mean value of $\beta_\mathrm{OA}$ while the dashed lines show the $1
\sigma$ uncertainty in $\overline{\beta_\mathrm{OA}}$.  The open
square shows the Beuermann et~al.~(\cite{BH99}) spectrophotometric
data.  The data are inconsistent with a constant $\beta_\mathrm{OA}$
at the 95\% confidence level (92\% confidence level if the
spectrophotometric datum is included).}
\label{FIGURE:grb990510_beta} 
\end{figure}

        We estimated the spectral index for the OA associated with
\object{GRB~990510} in the same way as we did for the OA associated
with \object{GRB~990123} (see Sect.~\ref{SECTION:grb990123_beta_OA}).
Fig.~\ref{FIGURE:grb990510_beta} shows the values of
$\beta_\mathrm{OA}$ that were determined in this manner.  The weighted
mean spectral index is $\beta_\mathrm{OA} = -0.531 \pm 0.019$ (SE)
with $\chi^2_{14} = 1.6598$, which suggests that we can reject the
hypothesis that the spectral index is constant with time at the 95\%
confidence level.  Our mean value for $\beta_\mathrm{OA}$ is
consistent (within $1.5 \sigma$) with the value obtained by Beuermann
et~al.~(\cite{BH99}) using long-slit spectrophotometry 3.9 days after
the burst.  They found $\beta_\mathrm{OA} = -0.55 \pm 0.10$ in the
interval 4900 {\AA} $\le \lambda \le$ 9000 {\AA}.
 
        Our result is consistent with the spectral index increasing
from $\beta_\mathrm{OA} \approx 0$ at the time of the burst to
$\beta_\mathrm{OA} \approx -0.6$ at the time of the break in the light
curve ($t_b = 0.82 \pm 0.28$ days after the burst).  After the break
the spectral index stays constant at $\beta_\mathrm{OA}$ of $-0.62 \pm
0.06$, which is consistent with the Beuermann et~al.~(\cite{BH99})
result.  We find $\beta_\mathrm{OA} = -1.29 \pm 0.23$ 3.6 days after
the burst.  This is inconsistent with the Beuermann
et~al.~(\cite{BH99}) data, but is consistent with the spectral index
continuing to increase after the break in the light curve occurred.
However, we believe that the direct measurement of Beuermann
et~al.~(\cite{BH99}) is more reliable than our indirect measurements
of $\beta_\mathrm{OA}$ so we conclude that it is unlikely that the
spectral index continues to increase after the break.

\section{Models\label{SECTION:models}} 
 
	The generic model for an OA associated with a GRB is a
fireball expanding relativistically into an ambient medium of number
density $n$, and decelerating as it sweeps up matter.  The shock
between the fireball and the medium accelerates electrons to
relativistic energies, and gives them a power-law distribution of
energies, $N(\gamma)\propto\gamma^{-p}$. The total energy in the
electrons is parameterized by $\epsilon_e$, the ratio between the
energy in electrons and the energy in nucleons.  Similarly, the
magnetic field strength is parameterized by $\epsilon_B$, the ratio of
the comoving field energy density and the post-shock nucleon energy
density (e.g.\ Wijers \& Galama~\cite{WG1998}).
 
    At a given instant in the fireball evolution the instantaneous
spectrum of the OA is determined by the local conditions in the shock
region.  In particular, the electron energy index $p$ determines the
spectral slope of the synchrotron emission $f_\nu \propto t^\alpha
\nu^{\beta}$, e.g., through the relation $\beta = -(p-1)/2$.
Following Sari et~al.~(\cite{SP98}), we have, for the fast cooling
regime,

\begin{equation} 
f_\nu = \left \{  
        \begin{array}{ll} 
                {(\nu/\nu_c)}^{-1/2}f_{\nu_m}, &        \nu_c < \nu < \nu_m,\\ 
                {(\nu_m/\nu_c)}^{-1/2}{(\nu/\nu_m)}^{-p/2}f_{\nu_m}, & \nu > \nu_c, 
        \end{array} 
        \right. 
\label{EQUATION:flux_fast} 
\end{equation}

\noindent
while in the slow cooling regime we have 

\begin{equation} 
f_\nu = \left \{  
        \begin{array}{ll} 
                {(\nu/\nu_m)}^{-(p-1)/2}f_{\nu_m}, &  \nu_m < \nu < \nu_c,\\ 
                {(\nu_c/\nu_m)}^{-(p-1)/2}{(\nu/\nu_c)}^{-p/2}f_{\nu_m}, & \nu_c < \nu. 
        \end{array} 
        \right.   
\label{EQUATION:flux_slow} 
\end{equation} 

\noindent
The synchrotron peak frequency, $\nu_m$, the flux at that frequency
$f_{\nu_m}$, and the frequency $\nu_c$ corresponding to the electron
Lorentz factor that separates the fast cooling electrons from the slow
cooling ones, all depend on the model parameters and the fireball
geometry (spherical or collimated). Explicit expression for these
cases can be found in Sari et~al.~(\cite{SP98}), Wijers \&
Galama~(\cite{WG1998}) (spherical), and Rhoads~(\cite{Rh99}) (jet). We
do not consider the spectrum below $\nu_c$ and $\nu_m$ in the fast and
slow cooling regimes, respectively, as this is not of interest in the
present discussion. The time dependence of $\nu_m$, $\nu_c$, and
$f_{\nu_m}$ is different in spherical and collimated models and has
been discussed by e.g.\ Rhoads~(\cite{Rh99}; see also Sari
et~al.~\cite{SPH99}).
 
        For a collimated outflow in an ambient medium of constant 
number density $n$, we have (Rhoads~\cite{Rh99})

\begin{equation} 
\nu_m = 3.66 \cdot 10^{16} x_p 
 \epsilon_e^2 \epsilon_B^{1/2} {\left(\frac{E_{52}}{\theta_0}\right)}^{1/2}  
       \frac{{(1+z)}^{1/2}}{t_d^{3/2} } \,\mathrm{Hz}, 
\label{EQUATION:nu_m} 
\end{equation}

\begin{equation} 
\nu_c = 1.04 \cdot 10^{11}\epsilon_B^{-3/2} E_{52}^{-2/3} n^{-5/6} \, \mathrm{Hz,~and} 
\label{EQUATION:nu_c} 
\end{equation}

\begin{equation} 
f_{\nu_m} = 1.8 \cdot 10^{-2} (1+z)\epsilon_B^{1/2} 
            \phi_p \left(\frac{E_{52}}{\theta_0^2}\right)  
                        n^{1/2}\frac{d_L}{4.82\,\mathrm{Gpc}}\,\mathrm{Jy}. 
\label{EQUATION:f_nu_m} 
\end{equation} 

\noindent
Here, $x_p$ is the dimensionless frequency at maximum flux and
$\phi_p$ is the dimensionless flux at that frequency (Wijers \&
Galama~\cite{WG1998}).  These are determined by the index $p$ of the
electron energy distribution.  The burst energy in units of $10^{52}$
erg is denoted by $E_{52}$, $\theta_0$ is the half angle of the
collimated jet, $t_d$ is the time in days, and $d_L$ is the luminosity
distance to the burst.
 
	To connect the optical observations in a given band pass
directly to the model parameters we rewrite, for example,
Eq.~(\ref{EQUATION:flux_slow}) as

\begin{equation} 
\nu^{(p-1)/2} f_\nu = \nu_m^{(p-1)/2} f_{\nu_m} 
\label{EQUATION:flux1} 
\end{equation} 

\noindent
for $\nu_m < \nu < \nu_c$ in the slow cooling regime, while for $\nu >
\nu_c$ we would have

\begin{equation} 
\nu^{p/2} f_\nu = {\left(\frac{\nu_m}{\nu_c}\right)}^{(p-1)/2} 
                  \nu_c^{p/2} f_{\nu_m}. 
\label{EQUATION:flux2} 
\end{equation} 

\noindent
Similar expression may be obtained in the fast cooling
regime. Combining $\nu_m$ and $f_{\nu_m}$ according to
Eq.~(\ref{EQUATION:flux1}), or $\nu_m$, $\nu_c$, and $f_{\nu_m}$
according to Eq.~(\ref{EQUATION:flux2}), depending on the frequency
range of interest, allows us to directly connect the observations at a
given frequency to the model parameters. The general formula that is
obtained this way is complicated but may be simplified in specific
cases (see Eq.~(\ref{EQUATION:combination})).
 
	The properties of the observed light curve of the OA, such as
the rate of decay $\alpha$, are primarily determined by the
hydrodynamic evolution of the fireball. In particular, spherical and
collimated bursts give rise to different forms of light curves.  The
adiabatic evolution of the Lorentz factor, $\Gamma$, of the fireball
is given by (Sari et~al.~\cite{SP98}; Sari \& Piran~\cite{SP99})

\begin{equation} 
\Gamma = 6.65 {\left(\frac{E_{52}}{n}\right)}^{1/8} t_d^{-3/8},
\label{EQUATION:gamma} 
\end{equation} 

\noindent
A spherical outflow generally results in a uniformly decaying light
curve that can have modest breaks which are due to the different time
evolution of the characteristic frequencies $\nu_m$ and $\nu_c$ (Sari
et~al.~\cite{SP98}). A pronounced break in the light curve may be
observed in a spherical outflow that becomes non-relativistic, but
requires the ambient medium to be denser, by several orders of
magnitude, than generally expected (Wijers et~al.~\cite{WRM1997}; Dai
\& Lu~\cite{DL99}). The steepening of the light curve in the case of a
spherical fireball entering a non-relativistic phase can be shown to
be $\Delta\alpha = (\alpha_1 - \alpha_2)= -(\alpha_1 + 3/5)$.
 
	Collimated outflows, on the other hand, give rise to a light
curve that resembles a broken power-law which becomes steeper between
several hours and several days after the gamma-ray event.  There are
two general possibilities.
 
\begin{enumerate} 
 
\item 
	There is a collimated outflow with a fixed opening angle
$\theta_0$. When $\Gamma$ decreases to $\Gamma \approx 1 / \theta_0$,
the light curve steepens.  This steepening occurs because after the
outflow has decelerated the observer only receives radiation emitted
within the collimated beam (M{\'e}sz{\'a}ros \& Rees~\cite{MR2000};
Kulkarni et~al.~\cite{KD99}).  The amount of steepening is
$\Delta\alpha = 3/4$.  This is a purely geometrical effect, so the
intrinsic properties of the fireball are not affected by the
transition to the $\Gamma < 1 / \theta_0$ regime.
 
\item 
	The collimated outflow may expand sideways as well as radially
(e.g.\ Rhoads~\cite{Rh99}; Panaitescu \&
M{\'e}sz{\'a}ros~\cite{PM98}). Initially, the relativistic beaming
dominates over the sideways expansion, but when sideways expansion
sets in, the radial expansion essentially comes to a halt, which
results in a steepening of the light curve.  Simple analytic estimates
show that this steepening is $\Delta\alpha = 1 - \alpha_1/3$
(Rhoads~\cite{Rh99}), while detailed numerical simulations show that
the steepening is expected to be more gradual both in time and slope
(Moderski et~al.~\cite{MS99}).
 
\end{enumerate} 
 
 
\section{Interpretation\label{SECTION:interpertation}} 
 
\subsection{GRB~990123} 
 
        The synchrotron peak frequency, $\nu_m$, was most likely in
the radio already on 1999 January 24.65 (Galama et~al.~\cite{GB99}),
implying that the electrons were already in the slow cooling
regime. The flux at frequency $\nu$ would then be given by
Eq.~(\ref{EQUATION:flux1}).  The isotropic energy of the burst,
$E_{52}$, that was estimated from the gamma-ray data, is $E_{52} = 338
\pm 1$ at the observed redshift of $z = 1.6$.

    The spectral index that we determined for \object{GRB~990123},
$\beta = -0.750 \pm 0.068$, corresponds to an electron index of $p =
2.500 \pm 0.136$.  There is no evidence for variations in $\beta$
across the break in the light curve, so we can conclude that $p$ is
constant during the observed evolution of the light curve. The
reliability of $\beta$ and $p$ obtained in this way is questionable,
however, as the effect of extinction on $\beta$ is unknown. The light
curve decay indices, $\alpha_1$ and $\alpha_2$, on the other hand, are
also a function of $p$, and are not affected by extinction.
Therefore, we have elected to use our fits to the measured light
curves to obtain the electron index, and use that to infer the
spectral index.
 
    In Table~\ref{TABLE:grb990123_parameters} we compare the electron
energy index that we calculate from the light curve indices using
three different models and assuming that the ambient medium has a
constant density.  During the initial phase, all models look spherical
to the observer, hence $p$ is the same for all of them. After the
break in the light curve, different models give different relationship
between the light curve decay rate and $p$, as indicated in
Table~\ref{TABLE:grb990123_parameters}. The last line in
Table~\ref{TABLE:grb990123_parameters} gives the size of the break
that is predicted by each model and should be compared to the observed
size of the break, $\Delta\alpha = 0.57 \pm 0.10$.  A spherically
expanding fireball entering a non-relativistic regime at the time of
light curve break gives the best overall agreement with the data (see
also Dai \& Lu~\cite{DL99}; Wang et~al.~\cite{WDL2000}).  That
interpretation, however, does have problems in accounting
simultaneously for the optical and radio properties of the burst
(Kulkarni et~al.~\cite{KF99}). It is worth noting that the sideways
expanding jet predicts a break in the light curve of
$\Delta\alpha=1.37$, which is more than twice the size of the observed
break.

\begin{table*} 
\begin{center} 
\caption{The electron energy index before, $p_1$, and after, $p_2$,
the break in the light curve, and the size of the break,
$\Delta\alpha$, inferred from the observations of \object{GRB~990123}
for the three models discussed in the text.  The observed values are
given for comparison.  The observed value of $p$ is computed from our
observed value of $\overline{\beta_\mathrm{OA}} = -0.750 \pm 0.068$.}
\smallskip 
\begin{tabular}{|c|r|r|r|r|} 
\hline 
    & \multicolumn{1}{c}{Spherical $+$ non-relativistic} 
        & \multicolumn{1}{c}{$\theta_0$ constant} 
        & \multicolumn{1}{c}{Sideways expansion} 
        & Observed \\ 
\hline 
$p_1$          &      $1-4\alpha_1/3 = 2.49 \pm 0.11$ & $1-4\alpha_1/3 = 2.49 \pm 0.11$ & $1-4\alpha_1/3 = 2.49 \pm 0.11$ & $2.50 \pm 0.14$ \\ 
$p_2$          & $(21-10\alpha_2)/15 = 2.53 \pm 0.04$ &  $-4\alpha_2/3 = 2.25 \pm 0.06$ &     $-\alpha_2 = 1.69 \pm 0.06$ & $2.50 \pm 0.14$ \\ 
$\Delta\alpha$ &    $-(\alpha_1+3/5) = 0.52 \pm 0.06$ &                           $3/4$ &  $1-\alpha_1/3 = 1.37 \pm 0.02$ & $0.57 \pm 0.10$ \\ 
\hline 
\end{tabular} 
\label{TABLE:grb990123_parameters} 
\end{center} 
\end{table*}

	Our preferred model in interpreting this burst is that of a
collimated outflow of a fixed opening angle. The main reason for this
choice is the model's ability to self-consistently account for the
observations at \emph{all} wavelengths.  If we accept the constant
$\theta_0$ model as the best description of \object{GRB~990123}, we
can deduce the remaining model parameters.  Evaluating
Eq.~(\ref{EQUATION:gamma}) at $t_d = t_b = 1.68$, and using
$E_{52}=338$, we find that the Lorentz factor of the fireball at the
time of the break is
 
\begin{equation} 
\Gamma = 5.47 {\left(\frac{E_{52}}{n}\right)}^{1/8} 
       \approx 11 n^{-1/8}. 
\end{equation} 
 
\noindent  
Therefore, interpreting the break in the light curve as a geometrical
effect, we find the opening angle of the collimated outflow to be
 
\begin{equation} 
\theta_0 \approx \frac{1}{\Gamma} 
         \approx \frac{1}{11} n^{1/8} 
         \approx 5^\circ n^{1/8}, 
\end{equation} 

\noindent
which enables us to eliminate the opening angle from the expressions
for $\nu_m$ and $f_{\nu_m}$
(Eqs.~(\ref{EQUATION:nu_m})~and~(\ref{EQUATION:f_nu_m})).  The opening
angle is weakly dependent on the number density of the medium where
the burst occurred.  Typical number densities in star-forming regions,
for example, can vary from $n \approx 1$ to $n \ga 1000$.  However,
this range of number densities corresponds to opening angles of
$5\degr \la \theta_0 \la 12\degr$, so the amount of collimation does
not depend strongly on the density of the ambient medium.
 
        The fireball evolution is independent of colour, so
observations in different bands passes should give identical results
when evaluating the left hand side of Eq.~(\ref{EQUATION:flux1}).
Using the data for Eq.~(\ref{EQUATION:broken_power_law}) (the broken
power law) in Table~\ref{TABLE:grb990123_fits} at the time of the
break in the corresponding light curve, averaging over all five band
passes, and rearranging, we obtain
 
\begin{equation} 
\epsilon_e^{p-1} \epsilon_B^{(p+1)/4}E_{52}^{(p+1)/4}n^{-(p-9)/32}  
    \approx 1.34\cdot 10^{-2}.
\label{EQUATION:combination} 
\end{equation} 
 
\noindent 
We have retained $p$ in the exponents of the model parameters to show
their explicit appearance, but used $p=2.37$ (the mean of $p_1$ and
$p_2$ for the constant $\theta_0$ model), to evaluate the right hand
side. Note that the above equation constrains four independent model
parameters, but the isotropic energy of the burst, $E_{52} = 338 \pm
1$, can be determined directly from observations, so only three
parameters remain undetermined.
 
        An additional expression can be obtained using the $2-10$ keV
$X$-ray data obtained on 1999 January 24.2(Heise et~al.~\cite{HD99}).
At that time the cooling frequency was less than the frequency of the
$X$-rays ($\nu > \nu_c$), so Eq.~(\ref{EQUATION:flux2}) applies.
Evaluating the left-hand side of Eq.~(\ref{EQUATION:flux2}) using the
$X$-ray observations we find
 
\begin{equation} 
\epsilon_e^{p-1} \epsilon_B^{(p-5)/4} n^{-(p-9)/32-5/6}E_{52}^{(3p-5)/12}  
        = 2089,
\end{equation} 

\noindent
and combining the optical and $X$-ray results gives
 
\begin{equation} 
\epsilon_B^{3/2} n^{5/6} \approx 1.33 \cdot 10^{-7}. 
\end{equation} 
 
\noindent 
This gives $\epsilon_B\approx 2.6\cdot 10^{-5}$ and $\epsilon_e\approx
0.79$ if $n=1$ cm$^{-3}$, or $\epsilon_B \approx 5.6 \cdot 10^{-7}$
and $\epsilon_e \approx 2.94$ if $n = 1000$ cm$^{-3}$.  An additional
relation is needed in order to determine $n$ independently.

	An alternative, but related interpretation is collimated
outflow in an ambient medium with a number density distribution of the
form $n(r) \propto r^{-\delta}$ (see Panaitescu et~al.~\cite{PMR98},
M{\'e}sz{\'a}ros et~al.~\cite{MRW98}).  Here, $\delta=0$ represents a
homogeneous distribution, while $\delta=2$ represents a distribution
characteristic of a pre-existing stellar wind. The light curve decay
rates will be steeper in a wind distribution than if the density is
constant, and the break in the light curve will be less pronounced.
In general $\Delta \alpha=(3-\delta)/(4-\delta)$.  This reproduces
$\Delta \alpha=3/4$ when $\delta=0$, as discussed above, but gives
$\Delta \alpha=0.5$ for a stellar wind density distribution ($\delta =
2$).

	Interpreting the burst using this model we obtain
$\delta=1.67\pm 0.54$, which is consistent with a stellar wind density
distribution.  This model adds $\delta$ as a new parameter, but
constrains $\beta$, since the spectral index is now fixed by the light
curve decay rates ($\alpha_1$ and $\alpha_2$) and $\delta$. We obtain
$\beta=-0.51\pm 0.14$, which is marginally consistent with the
spectral index inferred from the data, resulting in $p=2.02\pm 0.28$.
The other model parameters, discussed above are similar to what is
found in the $\delta = 0$ case.

\subsection{GRB~990510} 
 
        \object{GRB~990510} was a strong burst, although none of its
observed properties were extraordinary in any way (Wijers
et~al~\cite{WVG1999}).  The lower limit to its redshift is $z = 1.619
\pm 0.002$ (Vreeswijk et~al.~\cite{VG1999}), and the isotropic energy
release was $E_{52} = 17 \pm 1$.  The afterglow was the first to show
a smooth break in the light curve rather than a sharp break at a fixed
point in observer time.
 
    The measured light curve decay index before the break, $\alpha_1 = 
-0.54 \pm 0.14$, is too small to be explained by adiabatic fireball 
scenarios as it leads to a value for the electron energy index, $p<2$, 
that is too small. It is, however, consistent with a radiative 
fireball (e.g.\ Sari et~al.~\cite{SP98}). In that case the light curve 
is predicted to decay as $t^{-4/7}$ with a corresponding spectral index 
of $\beta = -1/2$. These values are both independent of $p$, and are in good 
agreement with the observed value of $\alpha_1$, and the value of 
$\beta$ that we inferred from the light curves. After the initial 
radiative stage, the fireball evolves to the more commonly observed 
adiabatic phase.
 
    In Table~\ref{TABLE:grb990510_parameters} we compare the electron
energy index, $p$, calculated from the observed value of $\alpha_1$
for the three models considered in the previous section.  As already
mentioned the initial slow decay is incompatible with the requirement
that the electron population have a finite energy, i.e.,~electron
energy indices of $p < 2$ correspond to infinite energy, and only the
sideways expanding jet model is consistent with the behavior of the
light curve at late time.  This model also gives good agreement for
the size of the break in the observed light curve, $\Delta \alpha =
1.44\pm0.17$.  The smooth break in the light curve (see also Stanek
et~al.~\cite{SGK1999}, Israel et~al.~(\cite{IMC1999}), and Harrison
et~al.~\cite{HB99}), further supports that interpretation (Moderski
et~al.~\cite{MS99}).  The late time spectral index, calculated from
$p_2$, is $\beta_\mathrm{OA} = -0.49 \pm 0.10$.  This is in good
agreement with the value determine from spectrophotometry
($\beta_\mathrm{OA} = -0.55 \pm 0.10$, Beuermann et~al.~\cite{BH99}),
and the mean value that we calculated from the light curve
($\overline{\beta_\mathrm{OA}} = -0.53 \pm 0.02$). Therefore, the
sideways expanding jet, with a radiative initial phase, is our
preferred model.
 
        Applying an analysis similar to the one that we did for
\object{GRB~990123}, we obtain an opening angle of $\theta_0 \approx
5\degr n^{1/8}$, which is consistent with the opening angle derived by
Harrison et~al.~(\cite{HB99}) for \object{GRB~990510} and the opening
angle that we derived for \object{GRB~990123}.  In principle it is
possible to derive $\epsilon_e$ and $\epsilon_B$ in the same manner
that we did for \object{GRB~990123}.  However, to do this we need an
observation of the flux at $\nu > \nu_c$, and, unfortunately, there
are no observations in the literature at these frequencies.
Therefore, we are not able to break the degeneracy between
$\epsilon_e$, $\epsilon_B$, and $n$ that is seen in
Eq.~(\ref{EQUATION:combination}).
 
\begin{table*} 
\begin{center} 
\caption{Predicted parameters for three models for gamma-ray 
bursts and the \object{GRB~990510} observations.} 
\smallskip 
\begin{tabular}{|c|r|r|r|r|} 
\hline 
    & \multicolumn{1}{c}{Spherical $+$ non-relativistic} 
    & \multicolumn{1}{c}{$\theta_0$ constant} 
    & \multicolumn{1}{c}{Sideways expansion} 
    & Observed\\ 
\hline  
$p_1$          &      $1-4\alpha_1/3 =  1.72 \pm 0.19$ & $1-4\alpha_1/3 = 1.72 \pm 0.19$ & $1-4\alpha_1/3 = 1.72 \pm 0.19$ & $1.35 \pm 0.14$ \\  
$p_2$          & $(21-10\alpha_2)/15 =  2.72 \pm 0.13$ &  $-4\alpha_2/3 = 2.64 \pm 0.25$ &     $-\alpha_2 = 1.98 \pm 0.19$ & $2.24 \pm 0.12$ \\  
$\Delta\alpha$ &    $-(\alpha_1+3/5) = -0.06 \pm 0.14$ &                           $3/4$ &  $1-\alpha_1/3 = 1.18 \pm 0.14$ & $1.44 \pm 0.24$ \\  
\hline 
\end{tabular} 
\label{TABLE:grb990510_parameters} 
\end{center} 
\end{table*}

 
\section{Conclusions\label{SECTION:conc}}

	It is possible to learn a great deal about the nature of a
GRB, and the environment where the burst occurred, from the optical
light curve of the burst's OA\@.  However, it is important that the
decaying light curve be well-sampled in order to accurately determine
the time and shape of the break, as well as the slope of the light
curve before and after the break.
 
	\object{GRB~990123} and \object{GRB~990510} are the first two
GRBs for which breaks have been observed in their optical light
curves.  We have collected all of the published photometry for these
two bursts and determined the time of the break and the slopes of
these light curves before and after the break using three different
fitting functions.  We also estimated the spectral indices of each OA
from near-simultaneous broad-band photometry.  These parameters were
used to constrain models for the nature of each burst.  Our results
suggest the following.
 
\begin{enumerate} 
 
\item Both \object{GRB~990123} and \object{GRB~990510} exhibit breaks
in their optical light curves at times between approximately one and
two days after the bursts.  There is no evidence that the time of the
break depends on frequency.  The break in \object{GRB~990123}'s
optical light curve appears to have occurred very fast, perhaps over a
period of less than five hours.  This is in sharp contrast to the
break in the optical light curve for \object{GRB~990510}, which
appears to have occurred slowly over approximately one day.
 
\item Unlike \object{GRB~000301C} (Jensen et~al.~\cite{JFG2000}) the
parameters of the light curve ($\alpha_1$, $\alpha_2$, and $t_b$) for
\object{GRB~990123} and \object{GRB~990510} are not strongly dependent
on the fitting function.  This means that our interpretations are not
strongly dependent on the type of fitting function used to derive the
slopes and break time.  However, the choice of fitting function can
influence the interpretation of the light curve for some GRBs.
 
\item Our favoured interpretation for the optical light curve of
\object{GRB~990123} is that the afterglow was collimated with a fixed
opening angle of $\theta_0 \approx 5\degr$.  Therefore, the observed
break in the light curve is a geometric effect that occurs when the
relativistic expansion of the fireball slows to $\Gamma \approx 1 /
\theta_0$.  This model reduces the energy released by the GRB by a
factor of $\approx 300$ to $E_{52} = 1.1$, which corresponds to the
conversion of $\approx 0.01$ Solar masses of material to gamma
radiation.  The magnetic field strength is $\epsilon_B \approx 2.6
\times 10^{-7}$ if the local number density is $\approx 1$ cm$^{-3}$
and $\epsilon_B \approx 5.6 \times 10^{-5}$ if the local number
density is $\approx 1000$ cm$^{-3}$.  This is well below the
equipartition energy.
 
\item The OA's light curve for \object{GRB~990510} is consistent with
a collimated outflow that is initially radiative. The transition to an
adiabatic phase, and the onset of a sideways expansion of the jet,
occurred near the time of the break. The opening angle was $\theta_0
\approx 5\degr$ at that time, the same as \object{GRB~990123}.  We are
unable to estimate reliably the magnetic field strength because no
measurements are available of the flux at frequencies above $\nu_c$,
the electron cooling frequency.

\end{enumerate} 

	We have investigated several models for \object{GRB~990123}
and \object{GRB~990510} and find that the OAs for these two bursts can
not be explained with a single model.  This suggests that the local
environment where a GRB occurs plays an important role in determining
the evolution of the fireball, and the observed light curve of the
OA\@.  We note, however, that the uncertainties in determining the
break time and slopes of the light curves make it difficult to
unambiguously determine the physics of the optical afterglow and its
environment.

 
\begin{acknowledgements} 
 
	This research was supported by the Danish Natural Science
Research Council (SNF), the Icelandic Research Council and the
University of Iceland Research Fund.  We wish to thank the referee for
several useful comments that allowed us to improve the manuscript.
 
\end{acknowledgements} 
 

\end{document}